\documentclass{nature}
\usepackage{amsmath}
\usepackage{amssymb}
\usepackage{graphicx}
\usepackage{float}
\usepackage{color}

\title{Extreme Ultraviolet Radiation With Coherence Time \\ Beyond 1 s}

\author{Craig Benko$^{1}$, Thomas K. Allison$^{1,2}$, Arman Cing\"oz$^{1,*}$, Linqiang Hua$^{1,3}$, Fran\c cois Labaye$^{1}$, \\Dylan C. Yost$^{1,**}$ \& Jun Ye$^{1}$}

\begin{document}

\maketitle

\begin{affiliations}
 \item JILA, National Institute of Standards and Technology and the University of Colorado, Boulder, CO 80309-0440
 \item Stony Brook University, Departments of Chemistry and Physics, Stony Brook, NY 11794-3400
 \item State Key Laboratory of Magnetic Resonance and Atomic and Molecular Physics, Wuhan Institute of Physics and Mathematics, Chinese Academy of Sciences, Wuhan 430071, People’s Republic of China
\end{affiliations}

\begin{abstract}

Many atomic and molecular systems of fundamental interest possess resonance frequencies in the extreme ultraviolet where laser technology is limited and radiation sources have traditionally lacked long-term phase coherence. Recent breakthroughs in XUV frequency comb technology have demonstrated spectroscopy with unprecedented resolution at the MHz-level, but even higher resolutions are desired for future applications in precision measurement. By characterizing heterodyne beats between two XUV comb sources, we demonstrate the capability for sub-Hz spectral resolution. This corresponds to coherence times $\mathbf{ > 1}$ s at photon energies up to 20 eV, more than 6 orders of magnitude longer than previously reported.
This work establishes the ability of creating highly phase stable radiation in the XUV with performance rivaling that of visible light. Further, by direct sampling of the phase of the XUV light originating from high harmonic generation, we demonstrate precise measurements of attosecond strong-field physics.
\end{abstract}

In the infrared, visible, and near-UV spectral regions, the precision and accuracy of frequency metrology\cite{udem2002, bloom2013} has vastly exceeded that of traditional spectroscopic methods. Two key enabling laser technologies driving these measurements are ultrastable lasers with long ($>$ 1s) coherence times\cite{Kessler2012, bishof2013} and optical frequency combs\cite{Ye_book2005}.  However, for wavelengths below approximately 200 nm, conventional light sources have short coherence times\cite{Attwood_book1999} and are not useful for frequency metrology. Future applications in precision measurement with highly-charged ions\cite{berengut2011}, helium\cite{eyler2007}, nuclear clocks\cite{campbell2012}, or hydrogen-like and helium-like ions\cite{herrmann2009} necessitate phase stable light in the XUV.  The workhorse of extreme ultraviolet (XUV) and soft x-ray science, synchrotron radiation\cite{Attwood_book1999}, is both temporally and spatially incoherent. New free electron laser sources possess a high degree of spatial coherence but due to their single-shot nature, coherence times are never longer than their (femtosecond) pulse durations\cite{ackermann2007,young2010,emma2010}. Conventional XUV lasers\cite{suckewer1990}, based on short-lived population inversions in highly ionized pulsed plasmas, are spatially coherent but also temporally incoherent on scales useful for XUV frequency metrology.

In the last decade, high-order harmonic generation (HHG) using femtosecond lasers has emerged as method that can support both spatial and temporal coherence in XUV\cite{Bellini_PRL1998} and use of HHG sources for high resolution spectroscopy in the XUV have recently been demonstrated at the MHz level\cite{cingoz2012,yost2011,kandula2010,morgenweg2014}. In this Article, we report the observation of coherence times greater than 1 second in the XUV measured via heterodyne beating between two XUV frequency comb sources based on intracavity HHG\cite{jones2005, gohle2005, Mills_JPhysB2012}. Similar to heterodyne interferometry with continuous-wave lasers, our frequency comb lasers in the near-infrared (NIR) are up-converted to the XUV and then combined to form the heterodyne beatnote.  The apparatus, schematically shown in Fig.(1a), is analogous to a Mach-Zender interferometer where each arm of the interferometer contains a cavity-enhanced HHG apparatus. The first beam splitter (the acousto-optic modulator) separates the NIR frequency comb and the second beam splitter (the beam combiner, Fig. (1d)) recombines the XUV frequency combs.  The heterodyne signal will provide information about the phase of the XUV light and its noise properties. We use the phase information of the heterodyne signal to probe attosecond time-scale, strong-field physics.

\subsection{XUV comb generation and XUV interferometer}
For our experiment, we pump two independent femtosecond enhancement cavities\cite{jones2002, jones2005, gohle2005} (fsECs) with a high-power Yb:fiber frequency comb outputting 120 fs pulses at 154 MHz with an average power of 80 W centered at 1070 nm\cite{ruehl2010}. To pump two fsECs, the frequency comb is split into two by an acousto-optic modulator (AOM) such that the two resulting combs have a carrier envelope offset frequency detuning of $\Delta f_{ceo} = $ 1 MHz. Each fsEC (XUV1 and XUV2 of Fig. (1a)) is actively stabilized using the Pound-Drever-Hall technique with piezo-electric transducers on cavity mirrors as actuators. The carrier envelope offset is not actively stabilized, but its passive stability is sufficient for maintaining good power enhancement. Each fsEC typically operates with $\sim$4.5 kilowatts of average power and peak intensity of $4\times10^{13}$ W cm$^{-2}$.  Xenon gas for HHG is injected at the cavity foci via quartz nozzles with $\sim 150\,\mu\text{m}$ aperture backed by $\sim$1 atmosphere of pressure. Harmonics are coupled out of the fsEC's in co-linear fashion using $330\,\mu\text{m}$ thick intracavity sapphire plates set at the Brewster angle for the fundamental radiation to limit intracavity loss.  The finesse of the fsEC was intentionally kept low to mitigate the nonlinearities of the Brewster plate and the plasma\cite{Mills_JPhysB2012, allison2011,carlson2011}, but enhancement factors of $\sim$200 were still obtained.  The two XUV beams are combined (more on this below) and detected using either an electron multiplier or a photomultiplier tube (PMT) with a phosphor screen.  A schematic of the optical layout is shown in Fig. (1a).  Since the pump frequency combs are offset by $\Delta f_{ceo}$, the resulting XUV frequency comb will have a relative detuning of $q \times \Delta f_{ceo}$ and heterodyne beatnotes can be observed at these frequencies so that a series of beatnotes appears in the RF output of the detector, effectively mapping the XUV spectrum of the harmonics to RF domain (Fig. (1c,1e)).

Splitting the NIR frequency comb beam at the beginning of the interferometer with the AOM is straightforward, but recombination of the XUV beams at the end of the interferometer is not. Since there are no transparent materials in the XUV, there are no standard beam-splitters available for recombination. Instead, we rely on a wavefront division scheme. A silicon wafer with a $100\,\mu\text{m}$ pyramidal aperture produced by KOH etching was used as the beam combiner (Fig. (1d)). The beam from XUV2 is focused through the aperture, while the reflected beam from XUV1 is much larger than the aperture. All the optics for the XUV light, including the beam combiner, were coated with boron carbide (B$_4$C) to enhance their reflectivity in the XUV\cite{vidal2008}. In the near field, there is no overlap between the two beams, but in the far field the beams interfere with a circular, ``bulls-eye", fringe pattern (see Fig. (1b)). The fringe pattern can be easily observed by setting $\Delta f_{ceo}$ = 0 and adjusting the relative path length of the interferometer so that pulses from XUV1 and XUV2 overlap in time at detector. To observe beat signals at finite $\Delta f_{ceo}$, the central portion of the fringe is selected with an aperture before detection. A key feature of this beam combination scheme is that the fringe period scales weakly as $\sqrt{\lambda}$ so that the interferometer can work well over a broad range of wavelengths, allowing us to simultaneously observe beats at the fundamental, the 17$^{th}$ harmonic, and all the harmonics in between.

Since fsEC's XUV1 and XUV2 are both pumped by a common NIR laser, the noise in the RF beatnote is immune to the common-mode frequency noise of the Yb:fiber laser. Thus, the apparatus directly measures noise from the HHG process or the cavity-plasma dynamics\cite{allison2011}. However, since the interferometer is not actively stabilized, there are small amounts of relative noise induced by vibrations in the optics, giving the two sources a non-zero relative linewidth that is technical in origin.

\subsection{Phase noise measurement and demonstration of long coherence times}
Here we show that high harmonic generation shares many common features with classical frequency multiplication of RF signals. In frequency multiplication, the power spectral density (PSD) of the phase noise $S_\phi(f)$ increases quadratically with the harmonic order\cite{walls1975}.  Even if the multiplication process is noiseless, any noise around the carrier will still increase with harmonic generation, thus setting a fundamental limit of how phase noise transforms under frequency multiplication and thus the achievable coherence level in the XUV (Methods).  The linewidth of a carrier can be related to $S_\phi(f)$ by the approximation\cite{hall1992},
\begin{align}
1 \,\text{rad}^2 \approx	\int^\infty_{f_{3dB}} S_\phi(f) \text{d}f.
\label{noise}
\end{align}
\noindent The limit of the integral $f_{3dB}$ is the point where approximately half of the power is in the carrier and half in the noise.  Therefore, the 3 dB linewidth of the carrier will increase quadratically with harmonic order.  If one is not careful to have a low-noise carrier before multiplication, the carrier will start to disappear and lead to the known phenomena of carrier collapse\cite{telle1996}.  Fig. (2) shows the full width at half max (FWHM) of each harmonic comb tooth plotted versus harmonic order.  A simple fit to FWHM = a q$^2$, where a is the scaling parameter, highlights the quadratic dependence of the linewidth.  The linewidth dependence on harmonic order was also independent of intracavity power.  This analysis shows that there is a fundamental scaling of phase noise from the generating laser to the resulting XUV light.  For example, assuming the best available continuous wave (CW) laser with a linewidth of 30 mHz\cite{bishof2013}, and that the fundamental comb will faithfully follow the phase of the CW laser\cite{schibli2008}, the linewidth at the 17$^{th}$ harmonic would be 8.7 Hz. Fortunately for HHG, we do not observe any additional linewidth broadening mechanisms other than the unavoidable classical frequency multiplication.

We show that the previously observed linewidth originated from the differential path noise and not HHG physics so that higher levels of coherence can be observed.  The differential path length noise in the interferometer can be removed by phase-stabilizing the two optical paths.  A small amount of pump light leaks through the XUV optics to the detection plane which is then picked off and detected on an photodetector to provide an error signal sensitive to interferometer fluctuations.  The error signal is filtered and used to apply feedback correction to the AOM frequency thus removing the relative noise (Methods).  With the interferometer phase locked, we can probe any noise processes intrinsic to intracavity HHG. A dramatic change in the linewidth of the XUV beatnotes is observed with a stable interferometer.  The unstabilized beatnote of the pump laser and the beatnote at the 17$^{th}$ harmonic are show in Fig. (3a,3c) with the corresponding stabilized case in Fig. (3b,3d) with a 250 mHz resolution bandwidth.  Similar resolution limited beats were observed on the 3$^{rd}$-19$^{th}$ harmonic.  The coherence of the pump lasers is faithfully transferred to the harmonics as seen by the resolution-limited beatnote of 62.5 mHz in Fig. (3f).  This is equivalent to a coherence time of 16 s or a stable phase relation maintained over $\sim 10^9$ consecutive pulses.  This is nearly seven orders of magnitude larger coherence times than ever reported in the XUV\cite{cingoz2012,yost2009,kandula2010}.  This establishes that our XUV comb system is extremely phase coherent and has the capability to support sub-Hz coherences in the XUV.  Furthermore, we demonstrate that the common-mode noise rejection of this measurement scheme is not particularly sensitive to the two XUV comb systems being identical by also observing coherent beats from harmonics generated from Xenon gas in one cavity and Krypton in the other (Fig. 3e).

\subsection{Application to attosecond strong-field physics}
The heterodyne beatnotes in the XUV provide unique and unprecedented access to the phase of XUV radiation.  We can use this to probe attosecond physics and measure the intensity dependent dipole phase\cite{lewenstein1995}.  This technique is a fundamentally different method for probing this phenomena than previous realizations and does not rely on extensive spatial and spectral filtering to observe interference between multiple quantum pathways\cite{yost2009,zair2008}.  This technique also does not rely on referencing the short to long trajectory contributions or vice versa\cite{corsi2006}.  It represents a complementary technique to RABBITT and related methods based on photoelectron spectroscopy\cite{mairesse2003, sansone2005, chang2011}, which seek to determine the time-delay (or equivalently phase shift) between adjacent harmonic orders at a given intensity. In this experiment, we isolate individual harmonics and measure precisely their phase shift as a function of intensity.  These phase shifts can be directly linked to temporal dynamics of the electron in the intense laser field\cite{mairesse2003}.  By putting amplitude modulation (AM) on one arm of the interferometer, we can measure the amount of induced phase modulation (PM) on the XUV light (Methods).  For this measurement, the peak intensity of the modulated cavity was $3.4\times10^{13} \text{ W cm}^{-2}$ with a 15\% amplitude modulation depth.   The intensity dependent dipole phase can be expressed as
\begin{align}
\phi = - \alpha_j \frac{U_p(I)}{\omega} = -\alpha_j\frac{I}{4\omega^3}
\label{idp}
\end{align}
\noindent with $I$ being the laser intensity, $U_p(I)$ being the pondermotive energy, $\omega$ being the laser frequency, and $\phi$ being the phase of the emitted XUV light. Due to our on axis spatial filtering needed to observe a high-contrast beatnote, we are primarily sensitive to the short trajectory.  The schematic of the measurement is shown in Fig. (4a).  The measurement required a two-step demodulation process to extract the amount of PM induced on the XUV light from the AM applied to the pump laser.  Great care was taken to ensure that the amount of PM on the XUV light was not induced by parasitic PM on the pump laser (Methods). The phase of the XUV light depends on both the intensity of the light and the particular quantum path the electron traverses (Fig. (4b)).  The result of the measurement is shown in Fig. (4c) where the $\alpha_j$ of Eq.~\ref{idp} is expressed in atomic units with the convention of Yost \emph{et al.}\cite{yost2009}. The shaded purple region corresponds to values predicted by the standard semi-classical model (SCM)\cite{lewenstein1995}.  The grey shaded region is the range of values predicted by different approximations to below threshold harmonics\cite{hostetter2010}.

Below-threshold, short trajectory harmonics do not originate from tunnel ionization as in the above-threshold case and are much more sensitive to the atomic potential and ionization dynamics\cite{hostetter2010}. Our measurement is able to discriminate between contributions of the standard SCM\cite{lewenstein1995} and a model that includes over-the-barrier (OTB) ionization\cite{hostetter2010} to confirm that the below-threshold harmonics mostly originate from the OTB ionization.  Further, our measurement for an above-threshold harmonic (15$^{th}$ and 17$^{th}$) agrees well with the predictions of the SCM\cite{lewenstein1995} and the below-threshold with predictions of Yost \emph{et al.}\cite{yost2009} and the theoretical framework of Hostetter \emph{et al.}\cite{hostetter2010}.  Further exploration of intensity dependent phases in atoms and molecules with comparison to calculations is the subject of future work. Our phase measurement technique was able to resolve phase shifts with uncertainties at the 10$^{-2}$ rad level, which corresponds to a time uncertainty of $\sim$ 6 as.  In contrast to typical experiments that utilize direct attosecond timing resolution\cite{chang2011}, we measure the attosecond electron dynamics imprinted on the phase of the emitted XUV light originating from HHG\cite{mairesse2003}.  With system improvements, it is feasible to extend this into the $\le 1$ as regime, rivaling the highest achievable temporal resolution of attosecond electron dynamics\cite{mairesse2003,krausz2009}. Thus, our apparatus provides direct, unambiguous access to the phase of XUV radiation and will prove to be a valuable tool for attosecond science and the dynamics of atoms and molecules in intense laser fields.

\subsection{Summary and future outlook}
We demonstrated that HHG is extremely phase coherent.  We have identified the primary noise requirements on the pump laser and shown that it is possible to support coherence times greater than 1 s in the XUV.  We have also developed an interferometer capable of operating from 1070 nm - 56 nm, but with different optics and a new beam combiner\cite{agaker2009}, extension to even shorter wavelengths is possible. Such an apparatus will be a vital tool for future work in dual-comb spectroscopy, Fourier transform spectroscopy\cite{keilmann2004}, high resolution molecular spectroscopy\cite{hinnen1995}, attosecond electron dynamics in intense laser fields\cite{shafir2013}, and HHG spectroscopy\cite{itani2004}.  We have successfully probed physics at attosecond timescales using the tools of frequency metrology to measure the intensity-dependent dipole phase.  Future work will require improved output coupling\cite{moll2006, pupeza2013, yost2008} and power-scaling\cite{yost2011, lee2011, carstens2014} schemes to extend the high level of phase coherence to shorter wavelengths possibly enabling spectroscopy of a nuclear isomer transition where highly phase-stable light will be needed for excitation.

\begin{methods}
\subsection{Acousto-optic Modulator and Beat Detection.}
In order to create a small frequency offset between the two XUV sources, we rely on an AOM to frequency shift the pump laser such that the relative detuning is 1 MHz.  Since there are no available AOMs at this frequency, we drive the AOM such that 1 MHz = $f_{rep}$ - $f_{AOM}$.  Therefore, we will observe two beats at frequencies less than $f_{rep}$, one at $f_{AOM}$ and the conjugate at $f_{beat} = f_{rep}-f_{AOM}$.  The conjugate beatnote is sensitive to any noise in $f_{rep}$.  To remove this, we derive $f_{AOM}$ by phase-locking a voltage-controlled oscillator to $f_{rep}$ detuned by 1 MHz such that $f_{AOM}$ = $f_{rep}$ - 1 MHz. This removes the dependence of $f_{rep}$ from the conjugate beatnote and puts it in $f_{AOM}$.  Therefore, the low frequency beatnote can be detected noise free where we have detectors of adequate bandwidth.

\subsection{Phase Noise.}
A signal oscillating at a frequency $\omega$ with phase modulation can be expressed as
\begin{align}
A = A_0 e^{-i \omega t-i\beta\text{sin}(\omega_m t)}
\label{pm}
\end{align}
\noindent where $\beta$ is the modulation depth and $\omega_m$ is the modulation frequency.  When $\beta$ is small, we can express the power in the first order modulation sideband relative to the carrier by $P_{SB}/P_{C} = \left( J_1(\beta)/J_0(\beta) \right)^2 \approx J_1(\beta)^2$, where $J_n$ are Bessel functions.  We can extend this to the case of general phase modulation and not at a particular discrete tone by $P_{SB}/P_{C}\approx (1/2)\Delta\phi_{rms}^2$. We can define the phase noise power spectral density as the noise around a carrier within a bandwidth $bw$ as
\begin{align}
S_\phi(f) \left[ \frac{\text{rad}^2}{\text{Hz}} \right] \equiv \frac{1}{2}\frac{\Delta\phi_{rms}^2(f)}{bw}.
\end{align}
\noindent By integrating $S_\phi(f)$, one arrives at Eq.~\ref{noise} and gets the approximate linewidth of a signal. The $f_{3dB}$ point is the cut-off of the integral where there is approximately half the power in the carrier and half in the noise.

\subsection{Phase-Stable Interferometer.}
The XUV light is very sensitive to path length fluctuations and any phase fluctuations in the driving laser.  Therefore, it is imperative to keep the interferometer stable in order to generate the highest levels of coherence in the XUV.  The passive stability is sufficient for sub-kHz levels of coherence.  To stabilize the interferometer, we use the small amount of pump laser light that co-propagates with the XUV light.  Since the pump light diverges more than the XUV light, it is simple to separate it out before the detection plane.  A small amount of light is picked off and sent to a photodetector where a 1 MHz beatnote is used to measure the phase fluctuations in the interferometer cause by mechanical noise on the mirrors.  The beatnote is mixed with an RF synthesizer to generate an error signal.  The error signal is filtered and the correction signal is applied to the 1 MHz reference frequency for the AOM.  Since the fluctuations of the interferometer are small,
the phase-lock loop that sets the AOM frequency can easily compensate for the small, necessary corrections.

\subsection{Measurement of Intensity Dependent Dipole Phase.}
The intensity dependent dipole phase is expressed in Eq.~\ref{idp}.  By trying to measure the intensity dependent phase that results from HHG, we are effectively measuring the AM-PM coupling with the AM being on the pump laser and the PM being on the XUV light.  We can mathematically describe a beatnote signal as
\begin{align}
S(t) = (1+\text{A}\,\text{Sin}(\Omega\,t+\phi_m))\,\text{Cos}(\omega\,t+\text{P}\,\text{Sin}(\Omega\,t+\phi_m)).
\label{eq2}
\end{align}
\noindent $\omega$ is the frequency of the beatnote and $\Omega$ is the frequency of the applied modulation and $\phi_m$ is its phase.  A is the amplitude modulation depth and P is the phase modulation depth.  Each beatnote is characterized by its own values for A and P.  To avoid confusion, the subscripts will refer to which signal it represents.  For example, A$_\text{IR}$ is for the amplitude modulation of pump laser and P$_\text{\text{q}}$ is the phase modulation of the q$^{th}$ harmonic.  Our task is to determine the values for A and P at the pump and the harmonics.  By taking the ratio of  P$_\text{q}$ to A$_\text{IR}$ and using proper units, we can extract the intensity dependent phase coefficient $\alpha_j$.

We need to extract the relevant parameters of Eq.~\ref{eq2} at both the harmonic of interest and the IR simultaneously.  To do this, we use a two step demodulation process.  By taking Eq.~\ref{eq2} and mixing it with a stable RF signal (LO1) at a frequency of $\omega$ and relative phase offset $\phi$, we get
\begin{align}
&S(t) \otimes V_1\text{Cos}(\omega\,t\,+\,\phi) = S_1(t) \\
S_1(t) \approx V_1(\text{Cos}(\phi)-\frac{\text{AP}}{2}&\,\text{Sin}(\phi)+\text{A}\,\text{Cos}(\phi)\text{Sin}(\Omega\,t+\phi_m)-\text{P}\,\text{Sin}(\phi)\,\text{Sin}(\Omega\,t+\phi_m)) .
\end{align}
\noindent We have ignored terms at $2\omega$.  We can further low pass the signals at $\Omega$ and obtain a ``DC" signal
\begin{align}
S_1(t) \overset{\text{Low Pass}}{\rightarrow} S_{DC} = V_1(\text{Cos}(\phi)-\frac{\text{AP}}{2}\,\text{Sin}(\phi)).
\label{DC}
\end{align}
\noindent Eq.~\ref{DC} will be one of our primary signals.  Note that the phase is set by the phase of LO1.  This also assumes that the phase of the XUV beatnote is stable. This is true when we phase stabilize our interferometer.

The signal $S_1$(t) contains terms that oscillate at the applied modulation frequency $\Omega$.  We can demodulate our signal once more using a lock-in amplifier at the correct phase $\phi_m$ and ignore terms at 2$\Omega$ to obtain
\begin{align}
S_1(t)\otimes V_2 \text{Sin}(\Omega t+\phi_m) = S_{LIA}\approx V_1V_2\,( \text{A}\,\text{Cos}(\phi)\, -\, \text{P}\,\text{Sin}(\phi))
\label{LIA}
\end{align}
\noindent $S_{LIA}$ is our second signal.  With Eq.~\ref{DC}, Eq.~\ref{LIA} and some independently measured parameters (for example, the AM-AM coupling), we can extract our parameters of interest.  To measure the A$_\text{IR}$-A$_\text{q}$ coupling, we can use our beatnote signals. Our XUV beatnotes are directly proportional to the amount of XUV power in each beam.  The amount of beatnote power can also be easily measured on an RF spectrum analyzer.  By changing the amount of power in one of the enhancement cavities and observing how the beatnote power changes, we can determine how much the XUV power must have changed for a given laser intensity change.  A$_{\text{q}}$ can be determined by
\begin{align}
\Delta\,\text{dB} &= 20\,\text{Log}_{10}(1/\sqrt{x})\\
\text{A}_\text{q}&= 1-x
\end{align}
\noindent with $\Delta$dB being the measured beatnote power.
By applying amplitude modulation to the pump laser on one arm of the interferometer, we can control A$_\text{IR}$ very well. It is also easily measured with a photodetector.  By varying the phase of LO1, we can measure $S_{DC}$ and $S_{LIA}$ (Eq.~\ref{DC} and Eq.~\ref{LIA}) simultaneously.  With the modulation (A,P) turned off,  $S_{DC}$ tells us the phase of the beat.  With the modulation on, the relative phase between $S_{DC}$ and $S_{LIA}$ can tell us the ratio of A/P.  Since A can be measured independently, we can extract the amount of phase modulation, P.  This procedure needs to be done with the IR signal and the XUV signal simultaneously to prevent any systematic errors.

The results of the measurements are shown in Fig.~(4c).  Each data point is the average of $\sim\,10$ runs of data where Eq.~\ref{DC} and Eq.~\ref{LIA} were measured while varying the phase $\phi$ of LO1 from 0 to 2$\pi$.  The measured values for P$_{\text{q}}$ were corrected for any parasitic P$_{\text{IR}}$ by P$_{\text{q}}\,\rightarrow\,$ P$_{\text{q}}$ - $\text{q}\,\times$P$_{\text{IR}}$.

\subsection{Signal Correction}
The signal correction of P$_{\text{q}}\,\rightarrow\,$ P$_{\text{q}}$ - $\text{q}\,\times$P$_{\text{IR}}$ is justified by experimental observations.  First, the quadratic scaling of the linewidth shows that the phase of the XUV light is directly linked to the phase of the NIR light.  Any phase shift $\Delta \phi$ on the NIR is $q \times \Delta \phi$ at the $q^{th}$ harmonic.  For the beatnote signals, this also implies that the phase modulation depth P$_{\text{q}}$ is related to the modulation depth in the NIR by P$_{\text{q}} = \text{q}\,\times$P$_{\text{IR}}$.  This relation is also justified by observing the beatnote signals in the presence of phase modulation on the NIR laser.  By measuring the amount of phase modulation on the NIR laser and the amount induced on the XUV beatnotes via the relation of the carrier to modulation sideband heights, we have verified the relation P$_{\text{q}} = \text{q}\,\times$P$_{\text{IR}}$.  Our lock-in detection methods also produce the same relation when addition phase modulation is placed on the NIR laser.

\subsection{Modulation Effects}
Modulating the intensity has notable effects on the neutral/plasma density ratio inside the enhancement cavity.  However, the modulation is slow at 2 kHz and the cavity feedback loop can easily follow it to maintain resonance.  The effect has been previously systematically studied\cite{allison2011}. Due to the neutral/plasma density changes, there is a small amount of phase modulation induced by amplitude modulation on the NIR laser.  This was verified by measuring the effect with gas present and absent.  This can easily corrupt the intensity dependent phase measurement and necessitates the correction described previously.  Further, any cavity oscillation due to bistability\cite{allison2011} can render the signals too noisy. Therefore, by careful measurement of the phase modulation on the NIR laser and the XUV beatnotes, we can extract the intensity dependent phase originating from HHG and not other macroscopic effects.
\end{methods}


\begin{addendum}
 \item We acknowledge technical contributions and collaborations from Axel Ruehl, Ingmar Hartl, and Martin Ferrman.  This work is supported by NIST, AFOSR, and the NSF Physics Frontier Center at JILA.
 \item[Competing Interests] The authors declare that they have no
competing financial interests.
 \item[Correspondence] Correspondence and requests for materials
should be addressed to J.Y.~(email: ye@jila.colorado.edu) or C.B.~(email: craig.benko@colorado.edu).\\
* Current Address: AOSense, Sunnyvale, CA  94085-2909 \\
** Current Address: Max-Planck-Institut f\"ur Quantenoptik, Hans-Kopfermann-Str.~1, 85748 Garching, Germany
 \item[Author contributions] All of the authors contributed to the design and performed the experiment.  All of the authors discussed the results and contributed to the final manuscript.
\end{addendum}

\newpage
\noindent \textbf{Fig. 1}

\begin{figure}
\centerline{\includegraphics[width= 6 in,height=6in]{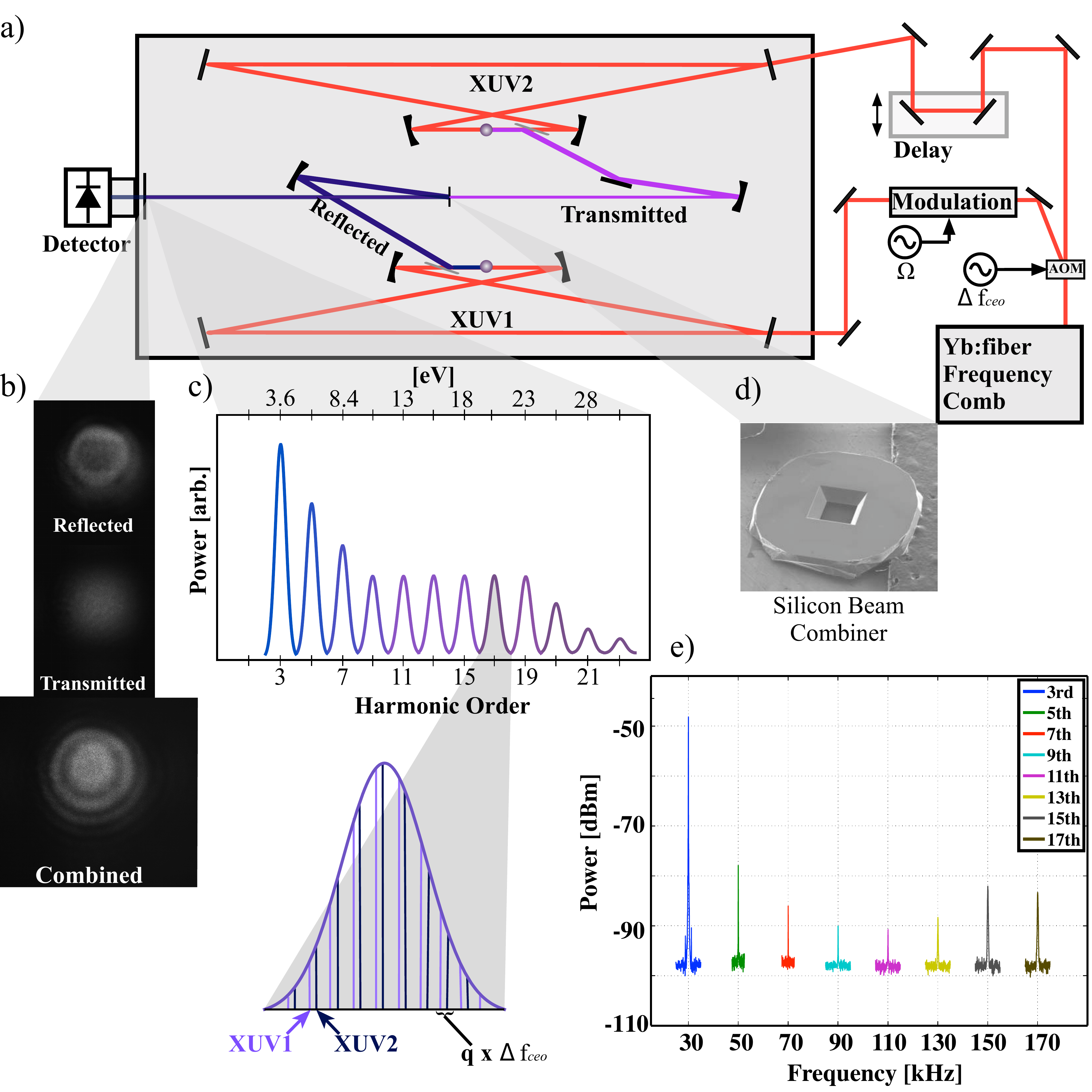}}
\caption{\textbf{Schematic of the experiment and harmonic spectra. } \textbf{a)} A high-powered frequency comb is split with and acousto-optic modulator (AOM) to create a  $f_{ceo}$ shift of 1 MHz.  The split beams are then coupled into independent femtosecond enhancement cavities (XUV1, XUV2) to reach intensities suitable for high harmonic generation.  The harmonic light is outcoupled with a Brewster plate and is steered to the beam combiner (shown in \textbf{d)}) with XUV optics.  \textbf{b)} Transmitted, reflected, and combined beam profiles.  The central interference fringe is selected with a spatial filter before detection.  \textbf{c)}  A schematic of a the harmonic spectrum.  Each harmonic order has frequency comb structure with teeth spaced by the laser repetition rate.  \textbf{e)} The measured radio-frequency beatnotes of the 3$^{rd}$-17$^{th}$ harmonic.  The beatnotes were mixed down to lower frequencies from q $\times$ 1 MHz for clarity.  The beatnotes are shown on a 30 Hz resolution bandwidth.}
\label{fig:fig1}
\end{figure}
\newpage
\noindent \textbf{Fig. 2}

\begin{figure}
\centerline{\includegraphics[width= 5.05 in,height=4.05in]{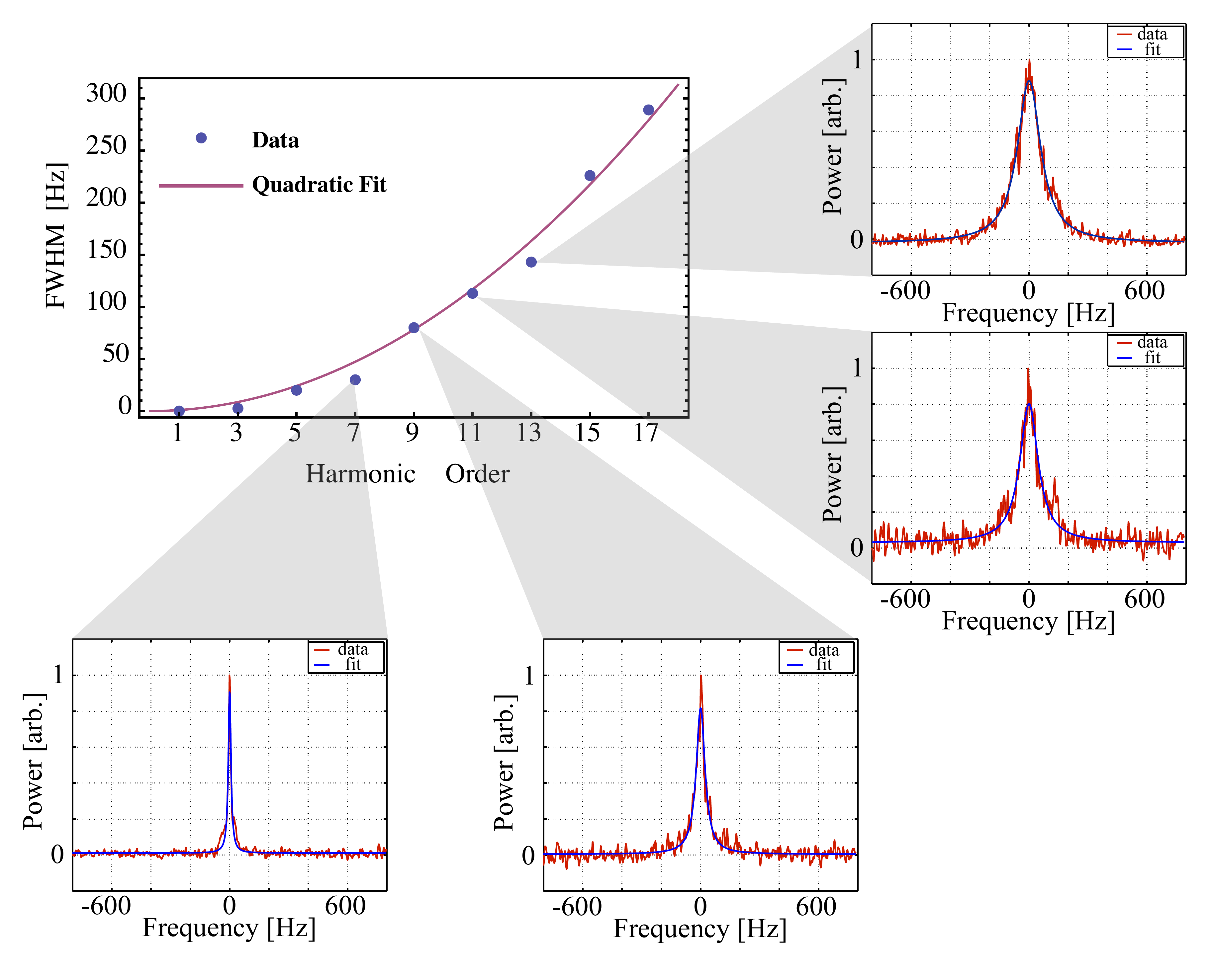}}
\caption{\textbf{Linewidth scaling of comb teether versus harmonic order.}  The fullwidth half max of the harmonic beatnotes are plotted versus harmonic order.  A clear quadratic dependence is shown by the fit.}
\label{fig:fig2}
\end{figure}
\newpage
\noindent \textbf{Fig. 3}

\begin{figure}
\centerline{\includegraphics[width= 4.375 in,height=5in]{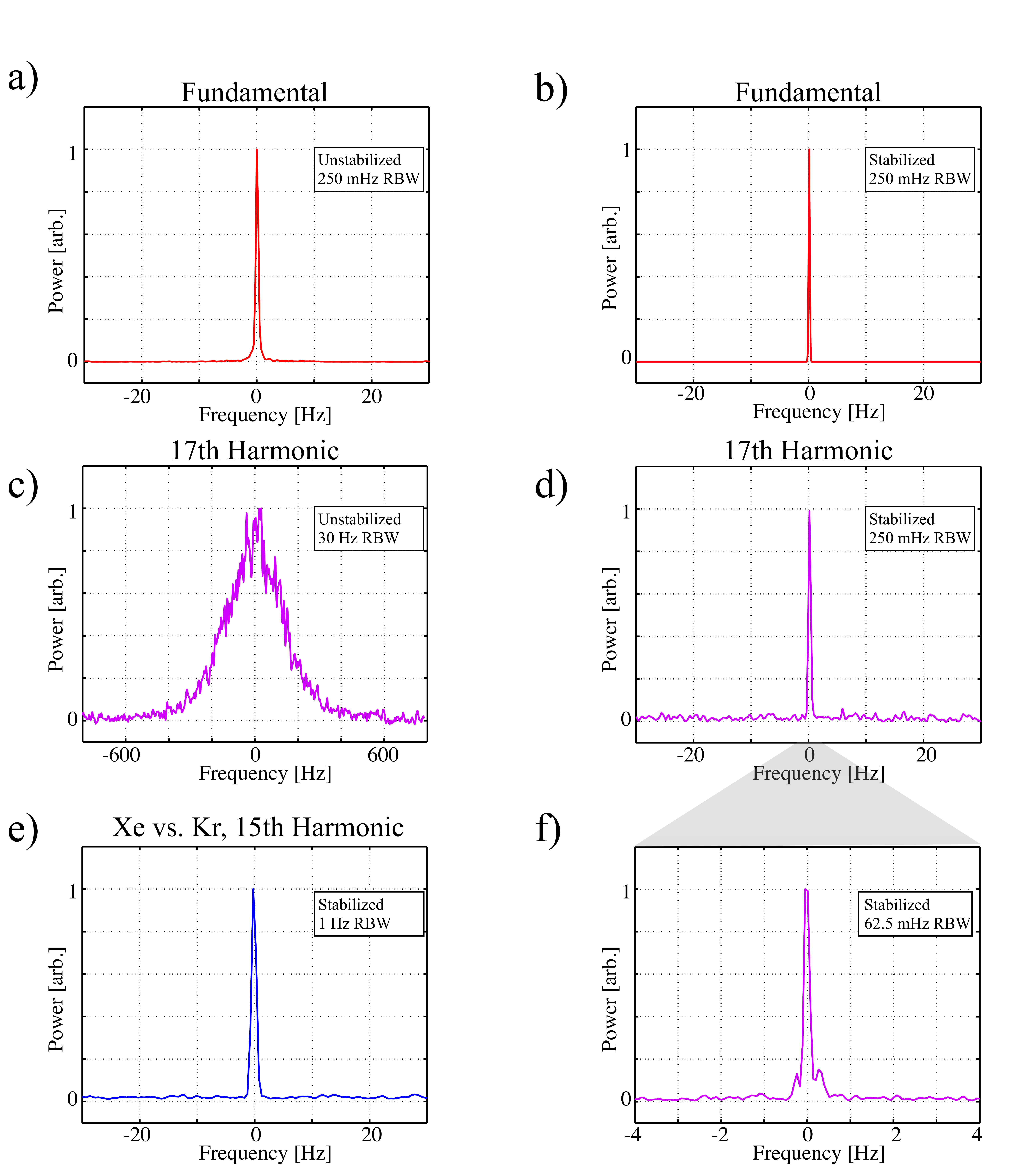}}
\caption{\textbf{Demonstration of sub-Hz coherence in the XUV.} \textbf{a,c)} Unstabilized beatnotes of the pump laser and the 17$^{th}$ harmonic.  \textbf{b,d)} Stabilized beatnotes of the pump laser and the 17$^{th}$ harmonic.  \textbf{f)}The resolution of the 17$^{th}$ harmonic can further be improved to a 62.5 mHz resolution bandwidth limited showing a coherence time of $>$ 16 s. \textbf{e)} The comparison of the two sources when one is injected with Krypton and the other with Xenon.  A 1 Hz resolution bandwidth limited signal is achieved.}
\label{fig:fig3}
\end{figure}
\newpage
\noindent \textbf{Fig. 4}

\begin{figure}
\centerline{\includegraphics[width= 5 in,height=3in]{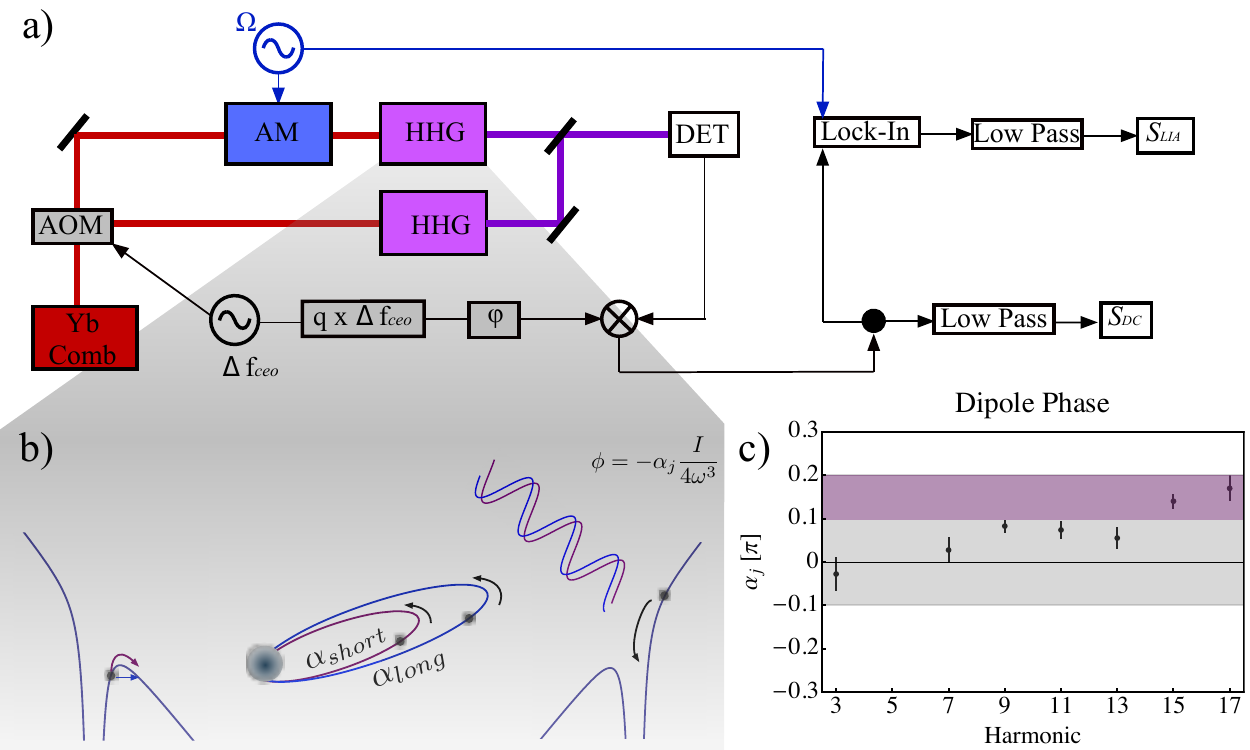}}
\caption{\textbf{Measurement of the intensity dependent dipole phase. }\textbf{a)} A schematic of the signal pathway.  A two step demodulation process is necessary to detect the phase modulation on the XUV light induce by amplitude modulation on the pump laser.  \textbf{b)} After an electron is ionized, it accelerates in the electric field.  Two dominant quantum trajectories (short and long) are important for the XUV emission where the phase of the light depends directly on the intensity of the electric field and its quantum path. \textbf{c)} Intensity dependent phase of the atomic dipole for the ``short'' trajectory of the 3$^{rd}$-17$^{th}$ harmonic.  The grey area is the predicted region where the below-threshold harmonic intensity dependent phase can fall.  The purple shaded region is where the above-threshold intensity dependent phase is predicted to lie.}
\label{fig:fig4}
\end{figure}
\newpage

\end{document}